\documentclass{desyproc}

\begin{document}
\title{Very-high-energy quasars hint at ALPs}

\author{{\slshape Marco Roncadelli$^1$, Giorgio Galanti$^{2}$, Fabrizio Tavecchio$^{3}$, Giacomo Bonnoli$^{4}$}\\[1ex]
$^1$INFN, Pavia, Italy\\
$^2$Universit\`a dell'Insubria, Como, Italy\\
$^3$Osservatorio di Brera-INAF, Milano, Italy\\
$^4$Osservatorio di Brera-INAF, Milano, Italy}

\contribID{familyname\_firstname}

\desyproc{DESY-PROC-2013-XX}
\acronym{Patras 2013} 
\doi  

\maketitle

\begin{abstract}
One of the mysteries of very-high-energy (VHE) astrophysics is the observation of flat spectrum radio quasars (FSRQs) above about $30 \, {\rm GeV}$, because at those energies their broad line region should prevent photons produced by the central engine to escape. Although a few astrophysical explanations have been put forward, they are totally {\it ad hoc}. We show that a natural explanation emerges within the conventional models of FSRQs provided that photon-ALP oscillations take place inside the source for the model parameters within an allowed range.
\end{abstract}

\section{Why do VHE quasars exist?}

With the advent of Imaging Atmospheric Cherenkov Telescopes (IACTs) H.E.S.S., MAGIC, VERITAS and CANGAROO III (the last one is not anymore operative) the VHE astrophysics has undergone a stunning development. Among the many discoveries, a remarkable one is that active galactic nuclei (AGN) emit photons up to energies of a few TeV. 

\begin{figure}[h]
\centerline{\includegraphics[width=0.20\textwidth]{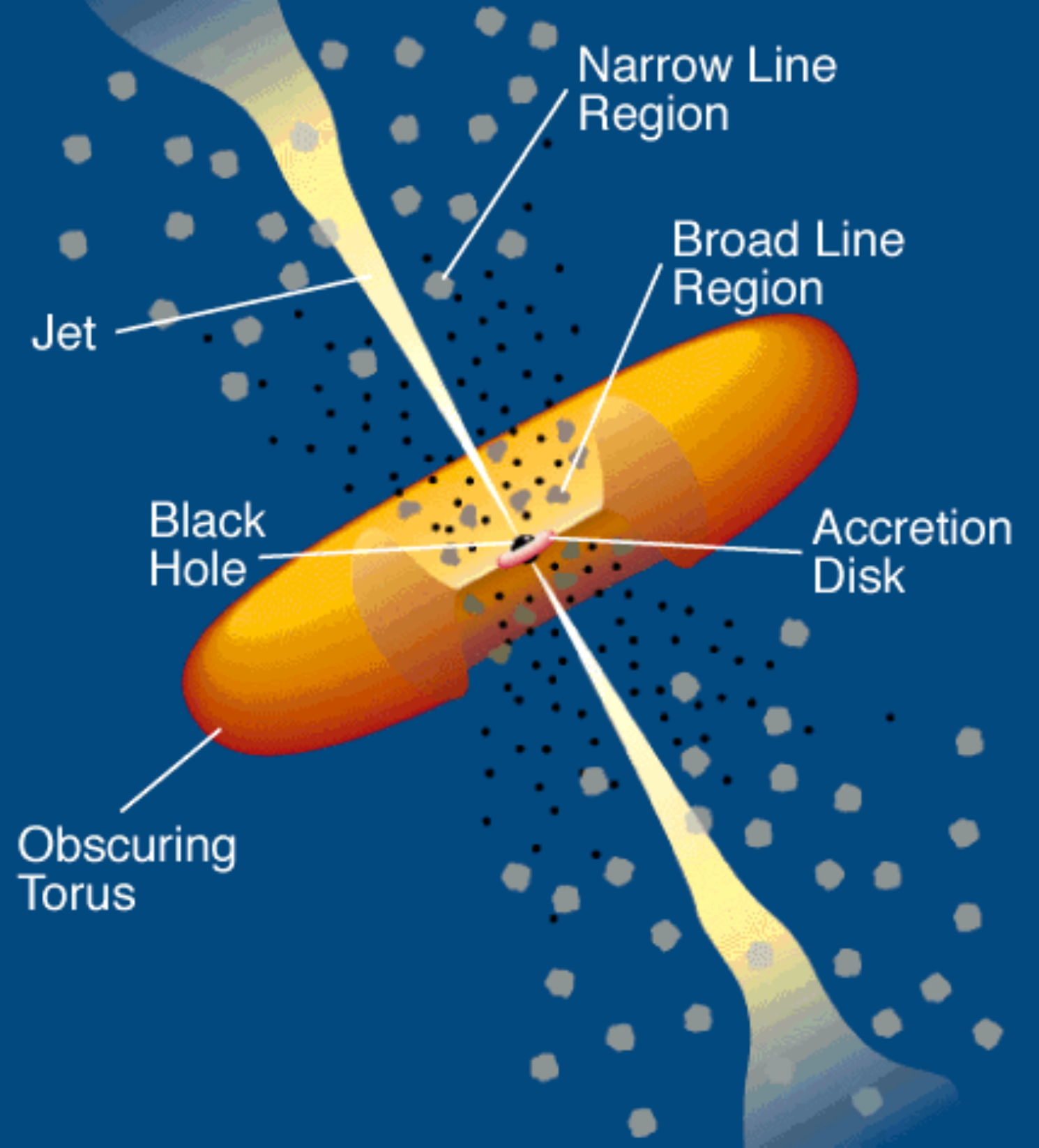}}
\caption{Schematic structure of an AGN.}\label{Fig:A}
\label{sec:figures}
\end{figure}

Before proceeding further, let us recall the basic properties of AGN. Basically, they are supermassive black holes (SMBHs) at the centre of elliptical galaxies. While observations have shown that SMBHs are hosted at the centre of both spiral and elliptical galaxies, in the former case they tend to be quiescent while in the latter one they are often active, namely they accrete matter which emits at all frequencies. Observations entail that such an emission is non-thermal, and two possible mechanisms have been envisaged. One is leptonic in nature, and usually called {\it synchro-self Compton} mechanism (SSC): owing to the presence of a magnetic field, relativistic electrons emit synchrotron radiation, and these photons acquire much larger energies by inverse Compton scattering off the parent electrons. In some cases also external electrons are responsible for the latter process. The resulting spectral energy distribution (SED) $\nu F_{\nu} \propto E^2 \, dN/dE$ has two peaks: the synchrotron one located somewhere from the IR to the X-ray band, while the inverse Compton peak lies in the $\gamma$-ray band around $50 \, {\rm GeV}$. The other emission mechanism is hadronic: the situation is the same for synchrotron emission, but the gamma peak is produced by hadronic collisions so that also neutrinos are emitted. In either case -- because the SMBHs is generally rotating -- it is surrounded by an accretion disk and the emission is strongly beamed in the direction perpendicular to the disk, which gives rise to two jets. As a consequence, they can be detected only when one of the jets points towards us. It has become customary to call {\it blazars} the AGN which are in such a configuration. This is schematically shown in Fig.~\ref{Fig:A}. 

\begin{figure}[h]
\centerline{\includegraphics[width=0.25\textwidth]{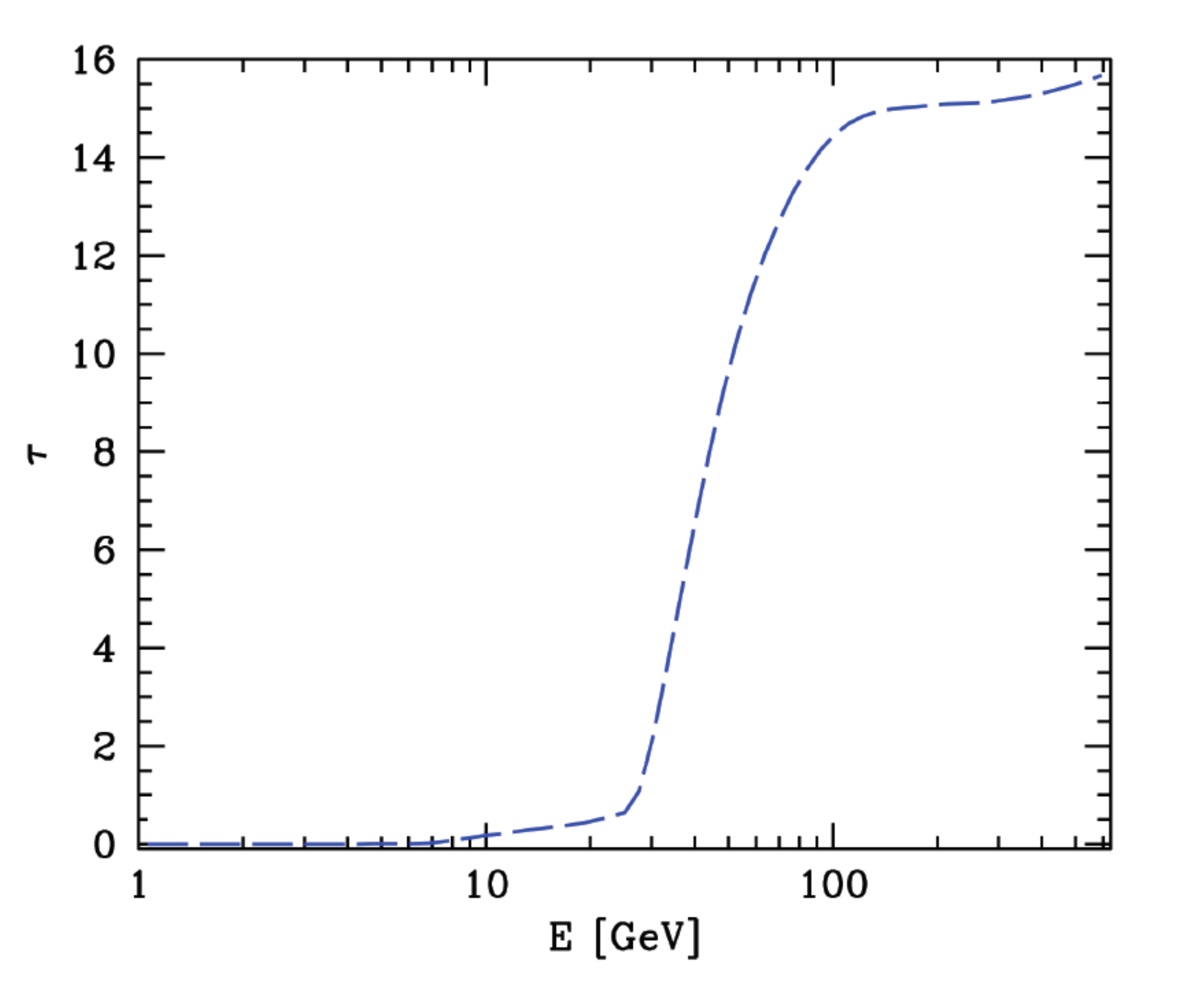}}
\caption{Energy behaviour of the optical depth in the BLR according to conventional models.}\label{Fig:B}
\label{sec:figures}
\end{figure}

As a matter of fact, blazars are commonly divided into two classes, depending on whether or not the broads line region (BLR) -- at about 1 pc from the centre -- is present or not. In the former case the blazar is called {\it flat spectrum radio quasar} (FSRQ), whereas in the latter case it is named BL LAC (this nomenclature is due to historical reasons). We remark that the BLR gives rise to broad optical lines, which where detected when the first quasars were discovered. Such a difference is very important for VHE astrophysics. In fact, in the BLR there is a huge density of ultraviolet photons so that the very-high-energy (VHE) photons ($E > 30 \, {\rm GeV}$) produced at the jet base undergo the process $\gamma \gamma \to e^+ e^-$, thereby disappearing from the spectrum. As a result, FSRQs should be invisible in the VHE band, as the optical depth depicted in Fig. \ref{Fig:B} eloquently shows.

However observations tell us that this is {\it not} true. For, at least 3 FSRQs have been detected by the IACTs in the energy range $100 \, {\rm GeV} - 1 \,  {\rm TeV}$: PKS 1222+216, 3C 279 and PKS 1510-089. And their fluxes are similar to those of the BL LACs! So, what is going on?

Actually, the most striking case is that of PKS 1222+216 which has been observed simultaneously by {\it Fermi}/LAT in the  band $0.3 - 3 \, {\rm GeV}$ and by MAGIC in the band $70 - 400 \, {\rm GeV}$. In addition, MAGIC has detected a flux doubling in about 10 minutes, which implies that the emitting region has size of about $10^{14} \, {\rm cm}$, but the observed flux is similar to that of a BL LAC. Thus, we have to face {\it two} problems at once!

Various astrophysical solutions have been proposed, but all of them are totally {\it ad hoc}, even because one has to suppose that a blob with size $10^{14} \, {\rm cm}$ at a distance of more than 1 pc from the centre exists with the luminosity of a whole BL LAC!

\section{Photon-ALP oscillations}

Nowadays our understanding of particle physics is based on the Standard Model (SM), which not only explains an enormous amount of data but has turned out to be fully correct with the discovery of the Higgs boson. Yet, it cannot be regarded as the ultimate theory. Apart from leaving the elementary particle masses and mixing angles as free parameters and ignoring gravity, it fails to provide any explanation for cold non-baryonic dark matter needed e.g. to explain cosmic structure formation, as well as for dark energy which presumably triggers the present accelerated cosmic expansion.  A generic prediction of many attempts towards the development of a  final theory -- like supersymmetric models, Kaluza-Klein theories and especially superstring theories -- is the existence of very light pseudo-scalar bosons characterized by a two-photon coupling, called {\it axion-like particles} (ALPs) due to their analogy with the axion.

As far as our analysis is concerned, the ALP Lagrangian has the form
\begin{equation}
\label{t1}
{\cal L}^0_{\rm ALP} = \frac{1}{2} \, \partial^{\mu} a \, \partial_{\mu} a - \frac{1}{2} \, m^2 \, a^2 + \frac{1}{M} \, {\bf E} \cdot {\bf B} \, a + \frac{2 \alpha^2}{45 m_e^4} \, \left[ \left({\bf E}^2 - {\bf B}^2 \right)^2 + 7 \left({\bf E} \cdot {\bf B} \right)^2 \right]
\end{equation}
where $a$ is the ALP field, the last term is the Heisenberg-Euler-Weisskopf effective Lagrangian accounting for the photon one-loop vacuum polarization in the presence of an external magnetic field ($\alpha$ is the fine-structure constant and $m_e$ is the electron mass), $m$ is the ALP mass and $M$ is a constant with the dimension of an energy. We stress that the parameters $m$ and $M$ are assumed to be {\it uncorrelated}, and it is merely supposed that $m < 1 \, {\rm eV}$ and $M \gg G_F^{- 1/2}$ with $G_F^{- 1/2} \simeq 250 \, {\rm GeV}$ denoting the Fermi scale. The only robust available bound on $M$ comes from the CAST experiment and reads $M > 10^{10} \, {\rm GeV}$.

In the presence of an external magnetic field, the two-photon coupling produces a mismatch between the interaction eigenstates and the propagation eigenstates, thereby giving rise to the phenomenon of photon-ALP oscillations. 

\section{A natural ALP-based explanation}

Our idea is remarkably simple. We assume that photons are produced by a standard emission model like the SSC at the jet base as in BL LACs, but that ALPs exist. Then photons can become mostly ALPs {\it before} reaching the BLR in the jet magnetic field. As a result, ALPs can go unimpeded through the BLR -- because $\sigma (a \gamma \to e^+ e^- ) \sim \alpha/M^2 < 10^{- 50} \, {\rm cm}^2$ -- and {\it outside} it they can reconvert into photons in the outer magnetic field. Because of lack of space, we cannot report the calculations which can be found in our original work~\cite{TRGG2012}, but we have found that the best choice to reduce the photon absorption by the BLR is $B = 0.2 \, {\rm G}$, $M = 7 \cdot 10^{10} \, {\rm GeV}$ and $m < 10^{- 9} \, {\rm eV}$. This choice leads to the results exhibited in Fig. \ref{Fig:C}.

\begin{figure}[h]
\begin{center}
\includegraphics[width=.35\textwidth]{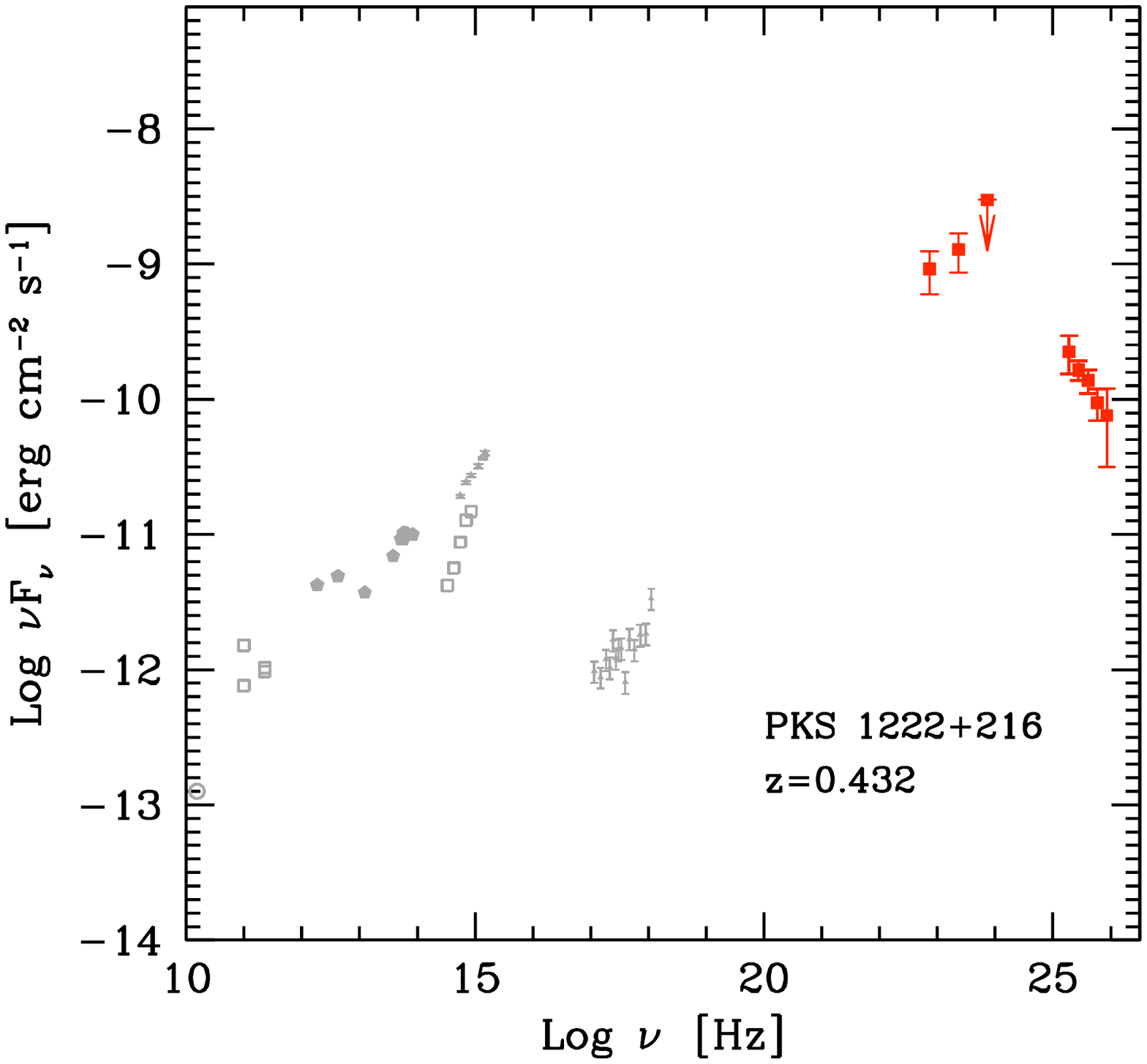}\includegraphics[width=.35\textwidth]{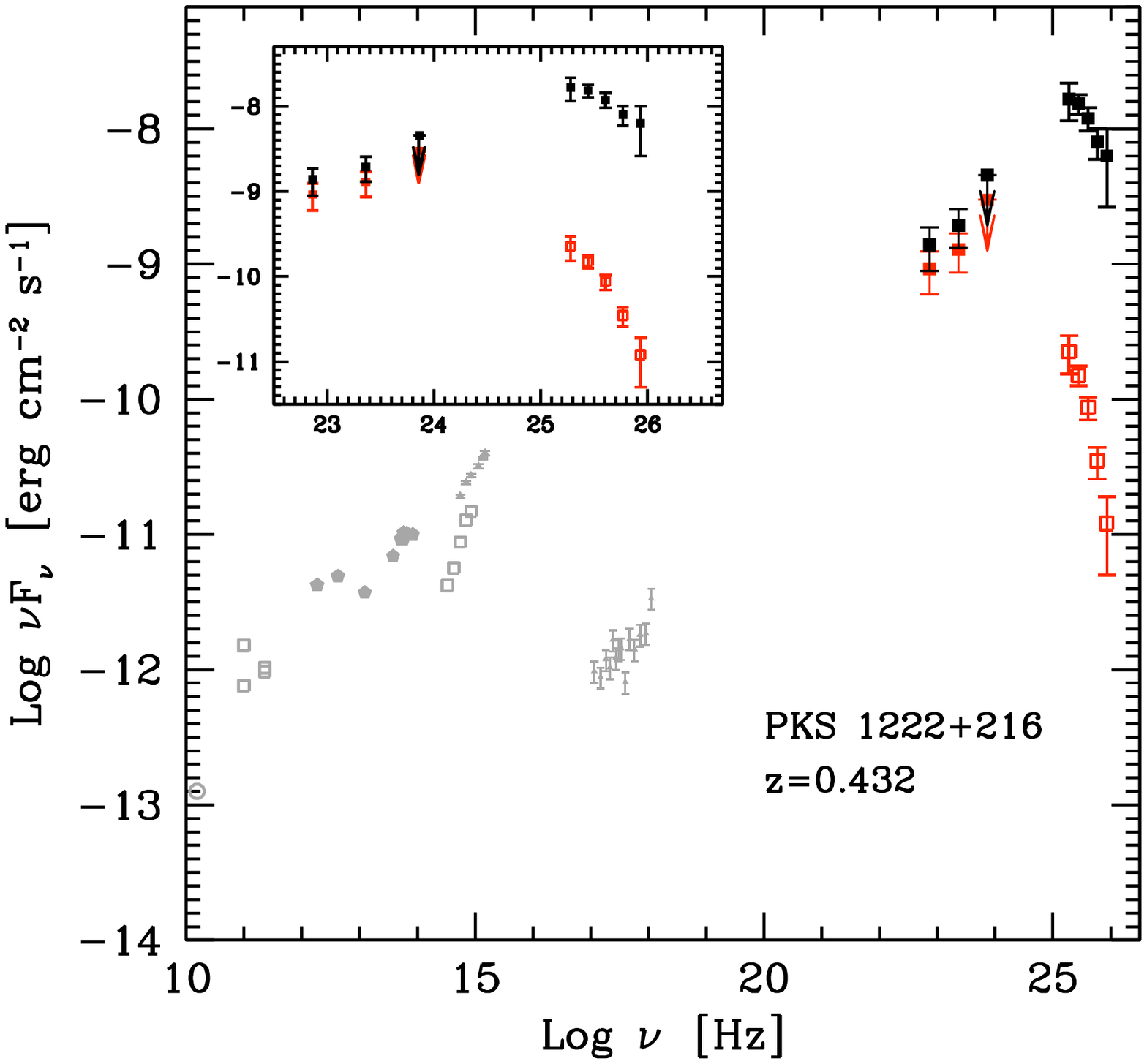}\includegraphics[width=.35\textwidth]{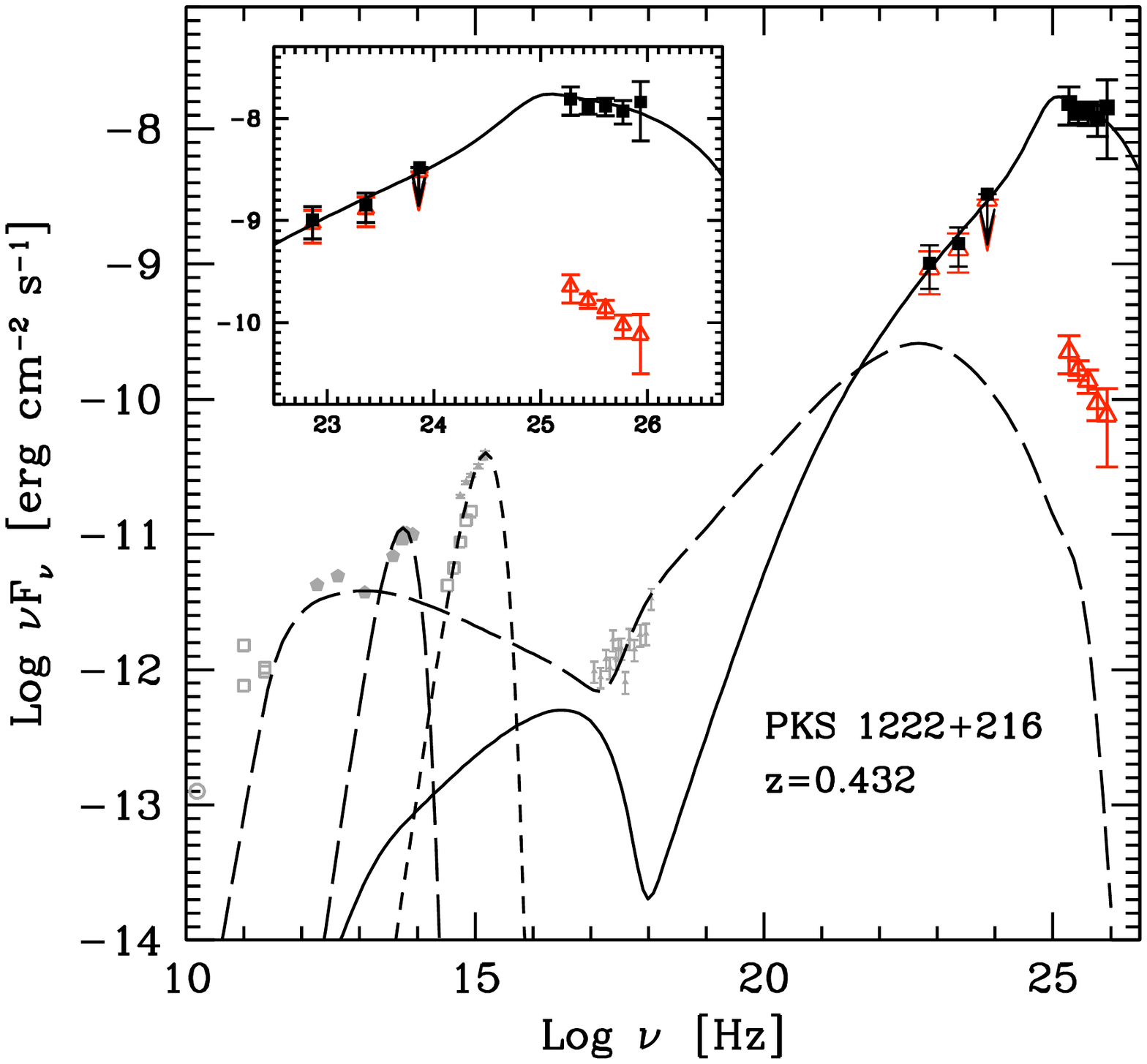}
\end{center}\label{Fig:C}
\caption{\label{Fig:C} 
Left panel: red triangles at high and VHE are the spectrum of PKS 1222+216 recorded by {\it Fermi}/LAT and the one detected by MAGIC but EBL-deabsorbed according to conventional physics. Central panel: red triangles are the same as before, while black squares represent the same data once further corrected for the photon-ALP oscillation effect. Right panel: red triangles and black squares are the same as before, whereas the solid black line is the SED of our model (the other points and broken lines should be presently ignored).}
\end{figure}

However, this is not enough. For, we have supposed that photons are produced by a standard emission mechanism. Moreover, PKS 1222+216 has been simultaneously observed by {\it Fermi}/LAT and MAGIC. So, we should pretend that the detected photons have a {\it standard} SED, namely that both data sets lie on the same inverse Compton peak. This requirement is {\it a priori} not guaranteed, since in the presence of absorption and one-loop QED effects the photon-ALP conversion probability is {\it energy-dependent}. Nevertheless, it turns out that a standard two-blob emission model with realistic values for the parameters yields the SED shown in the right panel of Fig.~\ref{Fig:C}. Hence, we see that the {\it Fermi}/LAT and MAGIC data indeed lie on the same inverse Compton peak.

Needless to say, our scenario naturally applies also to the other FSRQs detected at VHE. 

It looks tantalizing that just the most favorable choice of the parameters mentioned above corresponds to the most favorable case for a large-scale magnetic field of $B = 0.7 \, {\rm nG}$ in the DARMA scenario that enlarges the ``$\gamma$-ray horizon" and provides a natural solution to the cosmic opacity problem (it requires $m < 1.7 \cdot 10^{- 10} \, {\rm eV}$ which is consistent with the present choice)~\cite{dgr}. Moreover, this kind of ALP is a good candidate for cold dark matter~\cite{acgjrr}.

Obviously the CTA will provide a crucial test of our model, but it is remarkable that a laboratory check will be performed with the planned upgrade of the photon regeneration experiment ALPS at DESY and with the next generation of solar axion detectors like IAXO.

\section{Acknowledgments}

M. R. acknowledges the INFN grant TAsP (ex FA51).

\begin{footnotesize}

\end{footnotesize}


\end{document}